\documentclass[twocolumn]{aastex7}

\usepackage{amsmath}
\usepackage{bm}
\usepackage{diagbox}\usepackage{tabularx}
\usepackage{makecell}
\usepackage{xcolor}
\usepackage{nicefrac}
\usepackage[version=4]{mhchem}

\graphicspath{{./}{Figures/}}

\defcitealias{2024arXiv240911151S}{SM25a}
\defcitealias{2025ApJ...990..117S}{SM25b}

\begin{document}
\title{ The Response of Planetary Atmospheres to the Impact of Icy Comets III: Impact Driven Atmospheric Escape}

\author{F. Sainsbury-Martinez}
\affiliation{School of Physics and Astronomy, University of Leeds, Leeds LS2 9JT, UK}
\email{f.sainsbury-martinez@leeds.ac.uk}
\author{G. Cooke}
\affiliation{Institute for Astronomy, University of Cambridge, Cambridge, UK}
\email{gjc53@cam.ac.uk}
\author{C. Walsh}
\affiliation{School of Physics and Astronomy, University of Leeds, Leeds LS2 9JT, UK}
\email{c.walsh@leeds.ac.uk}

\begin{abstract} 
In an Earth-analogue atmosphere, water vapour is a key carrier of hydrogen in the lower atmosphere. The vertical transport of water to above the tropopause is one of the primary control valves on the atmospheric hydrogen escape rate. On the Earth, this escape is limited by transport though the tropospheric cold trap where water vapour condenses. However, on a tidally-locked exoplanet, the strong day-night temperature gradient drives a global-scale circulation. This circulation could rapidly transport water through the cold trap, potentially increasing hydrogen escape and impacting the composition of potentially habitable worlds.
We couple cometary impact and planetary atmospheric models to simulate water-depositing impacts with both a tidally-locked and Earth-analogue atmosphere and quantify how atmospheric circulations transport water from the impact site to high altitudes where it can potentially drive escape. 
The global nature of the atmospheric circulations on a tidally-locked world enhances hydrogen escape, with both our unimpacted tidally-locked and Earth-analogue atmospheres exhibiting similar mass loss rates despite the tidally-locked atmosphere being both cooler and drier near the surface. When considering the effects of a cometary impact, we find an order of magnitude difference in peak escape rates between impacts on the day-side ($\Phi_{\mathrm{escape}}=1.33\times10^{10}\,\mathrm{mol\,mth^{-1}}$) and night-side ($\Phi_{\mathrm{escape}}=1.51\times10^{9}\,\mathrm{mol\,mth^{-1}}$) of a tidally-locked atmosphere, with the latter being of the same order of magnitude as the peak escape rate found for an impact with an Earth-analogue atmosphere ($\Phi_{\mathrm{escape}}=2.7\times10^{9}\,\mathrm{mol\,mth^{-1}}$). Our results show the importance of understanding the underlying atmospheric circulations when investigating processes, such as hydrogen escape, which depend upon the vertical advective mixing and transport.
\end{abstract}


\section{Introduction} \label{sec:introduction}

Material delivery by cometary and asteroidal impacts has long been invoked as a key mechanism which shaped the early-Earth's composition and atmosphere.
For instance, numerical models of the migration of small bodies from the outer solar system, where ices can form (see the review by \citealt{snodgrass2017}), towards the inner solar system suggests that the total mass of water that they can deliver to the Earth is on the order of the mass of the Earth's oceans (\citealt{2003EM&P...92...89I,2004AdSpR..33.1524I,2014Icar..239...74O,2018SoSyR..52..392M,2023PhyU...66....2M}). This may explain why the Earth's surface is relatively water rich despite its interior being dry (\citealt{2020MNRAS.499.5334S}). 

It has been argued that impacts may play a similar role in other planetary systems, delivering complex organic (prebiotic) molecules (\citealt{2023RSPSA.47930434A}) and/or volatiles (\citealt{2020A&A...638A..50F,2024ApJ...966...39S}), enriching the secondary atmospheres of potentially habitable exoplanets. 
Recently, \citet{2024arXiv240911151S} (henceforth \citetalias{2024arXiv240911151S}) and \citet{2025ApJ...990..117S} (henceforth \citetalias{2025ApJ...990..117S}) explored the effects of an icy cometary impact on the climate and composition of both a tidally-locked and exo-Earth-analogue  exoplanetary atmosphere. These models revealed the importance of a planet's atmospheric circulation on the climate response to a cometary impact. On a tidally-locked world, the fixed day-night insolation drives a global scale circulation which is highly efficient at carrying material aloft. Whereas on an Earth-analogue planet with a diurnal cycle, mixing is strongest near the surface, and vertical mixing is delayed/weakened relative to a similar tidally-locked world. However, in either case, these atmospheric circulations carry at least a fraction of the deposited water aloft into the lower pressure (high altitude) regions of the atmosphere. Here, during the day, or on the day-side, the water is exposed to significant UV irradiation, driving photodissociation which frees both hydrogen and oxygen. The heavier oxygen molecules go on to react with other atmospheric constituents, generally increasing the abundance of oxygen rich molecules (\citetalias{2024arXiv240911151S}/\citetalias{2025ApJ...990..117S}).

However, hydrogen, as the lightest of the elements, is most easily lost to space. There are several processes which can drive this loss, including; Jean's escape, hydrodynamic flows, sputtering, photochemical escape, and charge exchange escape (see the reviews by \citealt{2018PhyU...61..217S} and \citealt{https://doi.org/10.1029/2019JA027639}). For all these mechanisms, the hydrogen escapes rates from the  top of the atmosphere are set by the vertical transport of hydrogen and hydrogen-rich, photochemically-sensitive, molecules to high altitudes. This is known as the diffusion-limited escape (\citealt{1974JAtS...31..305H,1977evat.book.....W,Catling_Kasting_2017}) regime. On the Earth there are two bottlenecks which limit this escape rate: the turbopause, where diffusion switches from being turbulent to molecular, and the cold trap at the tropopause, which limits the upwards advection of \ce{H2O} vapour by acting as a condensation trap. On the Earth, the cold trap is partly responsible for a decrease in water vapour concentration from an average of $\sim 1000$ ppmv in the troposphere to $\sim 5$ ppmv at the top of the stratosphere (\citealt{2020JGRA..12527639G}). As water is the primary carrier of hydrogen at these pressures, this will correspond to a similar multi order-of-magnitude decrease in the hydrogen escape rate. \\
However the situation is likely to be different on an exoplanet and/or after an impact. For example, on a tidally-locked world, the constant illumination of the day-side leads to a global scale atmospheric circulation which may help to transport super-saturated water through the cold trap and into the outer atmosphere (a similar effect can occur on Mars - \citealt{doi:10.1126/science.1207957}). Further, after an impact, deposited water can act as an opacity source, absorbing incoming radiation and thus changing the local temperature of the atmosphere, potentially influencing the effect of the cold trap. Finally, the large amount of water deposited at a single location will increase the molecular abundance gradient, promoting both horizontal and vertical diffusion. Taken together with atmospheric circulations, these effects drove the significant increase in the abundance of high altitude water and hydrogen found by \citetalias{2024arXiv240911151S} and \citetalias{2025ApJ...990..117S}. 

In this work, we build upon that of \citet{egusphere-2025-1133} who studied how hydrogen escape rates would have changed throughout the Earth's oxygenation history using Earth-atmosphere models calculated using CESM (the Community Earth System Model). Here, we quantify atmospheric escape rates after an icy cometary impact for two planetary scenarios: a tidally-locked Earth-like planet modelled after TRAPPIST-1e (\citetalias{2024arXiv240911151S}) and an Exo-Earth-analogue with a diurnal cycle (\citetalias{2025ApJ...990..117S}). Furthermore, due to the important role of the global overturning circulation in the vertical advection of impact-delivered water in a tidally-locked atmosphere (\citetalias{2024arXiv240911151S}), as well as the role of land-masses in driving asymmetries in this global wind (\citealt{anand2024,sainsbury2024b}), we will investigate how the hydrogen escape rate changes with impact location on our tidally-locked world. Note that, for the sake of both simplicity and comparisons with an epoch of the Earth's climate history for which the diffusion-limited hydrogen escape rate is relatively well constrained, we focus on impacts with an atmosphere that resembles the modern Earth. Future studies will investigate the effects of cometary impacts on the composition, and hydrogen escape rate, of Earth-analogue and tidally-locked atmospheres at earlier stages in a Earth-like-planets oxygenation history.

In \autoref{sec:method} we introduce our coupled impact and climate model, and give an overview of how we calculate the diffusion-limited escape rate. In \autoref{sec:results}, we quantify the effect of a cometary-impact on hydrogen escape rates for both Earth-analogue and tidally-locked exoplanets, focusing on the role that vertical transport plays in delivering hydrogen and hydrogen-rich molecules to high altitudes where escape primarily occurs. In \autoref{sec:concluding_remarks} we finish with some concluding remarks, discussing the implications of our results for the oxygenation and metallicity of potentially habitable atmospheres, as well as the overall importance of understanding the underlying circulation regime of a planet when quantifying atmospheric escape rates and the climatic response to external material delivery. 

\section{Method} \label{sec:method}

\subsection{Coupled Impact/Climate Model} \label{sub:coupled_impact_climate_model}

\begin{deluxetable}{lcc}

\tablecaption{Parameters of the impacting pure water ice comet considered in this work. \label{tab:comet_parameters}}
\tablehead{Parameter & Value & Unit}
\startdata
Radius $R$ & $2.5$ & km\\
Density $\rho_c$ & 1 & g cm$^{-3}$ \\
Mass $M$ & $6.5\times10^{16}$ & g \\
Initial Velocity $V$  & 10 & km s$^{-1}$  \\
Heat Transfer Coefficient $C_{H}$  & 0.5 & -\\
Drag Coefficient $C_{D}$ & 0.5 & - \\
Latent Heat of Ablation  $Q$ & $2.5\times10^{10}$ & erg g$^{-1}$ \\
Tensile Strength $\sigma_{T}$ & $4\times10^{6}$ & erg cm$^{-2}$
\enddata
\end{deluxetable}

We couple the parametrised impact model of \citetalias{2024arXiv240911151S} with a modified\footnote{\url{gitlab.com/leeds_work/cesm_comet}} version of the Earth-System-Model WACCM6/CESM2 which includes functions to describe thermal energy deposition and tidally-locked insolation (\citealt{2023ApJ...959...45C}). WACCM6, the Whole Atmosphere Community Climate Model is a well documented (\citealt{https://doi.org/10.1029/2019JD030943}), high-top (extending from the surface to $10^{-8}$ bar, corresponding to altitudes of $h\sim140$ km for the Earth and $h\sim160$ km for a tidally-locked world), configuration of the open-source Community Earth-System Model (CESM2). It includes a modern, Earth-like, land-ocean distribution (including orography) that is coupled to the atmosphere, including coupled dynamics, (photo)chemistry, and radiative transport.It was modified by \citet{2023ApJ...959...45C} to account for the effects of synchronous rotation. And an older version of the model (based on CESM 1.2.1), referred to as ExoCAM, has been benchmarked against other exoplanetary GCMs as part of the TRAPPIST-1 Habitable Atmosphere Intercomparison (THAI) project (\citealt{2022PSJ.....3..211T,2022PSJ.....3..212S}).
Here we consider two initial climate states: a tidally-locked planet with planetary/orbital parameters matching those of TRAPPIST-1e and a pre-industrial Earth-like \ce{N2-O2} atmosphere (\autoref{ssub:trappist}), and an exo-Earth-analogue atmosphere (\autoref{ssub:exo_earth_analogue}).

Our cometary impact model simulates the passage of a relatively low tensile strength comet ($\sigma_{T}=4\times10^{6}$ erg cm$^{-2}$) though the atmosphere of a terrestrial (exo)planet. At high altitudes, where the atmospheric density is low, the comet slows due to atmospheric drag which drives surface ablation and hence thermally-driven mass deposition. This drag (and hence ablation) increases in strength as the comet travels to lower altitudes, until eventually the drag and associated stresses exceed the tensile strength of the comet. At this point we consider breakup to have occurred and distribute any remaining mass over a pressure scale height to mimic the resulting deposition profile due to the remaining inertia of the impacting material.  A full overview of this model, including the mass and energy deposition profile for an icy cometary impact with the sub-stellar point of our TRAPPIST-1e atmospheric model (\autoref{ssub:trappist}) can be found in section 2.1 (figure 1) of \citetalias{2024arXiv240911151S}. Similarly, the mass and energy deposition profile for a icy cometary impact over the Pacific Ocean of an exo-Earth-analogue can be found in section 2.1 (figure 1) of \citetalias{2025ApJ...990..117S}. Parameters of the impacting pure water comet considered here can be found in \autoref{tab:comet_parameters}. 

\subsubsection{TRAPPIST-1e} \label{ssub:trappist}
\begin{figure*}[tp] %
\begin{centering}
\includegraphics[width=0.7\textwidth]{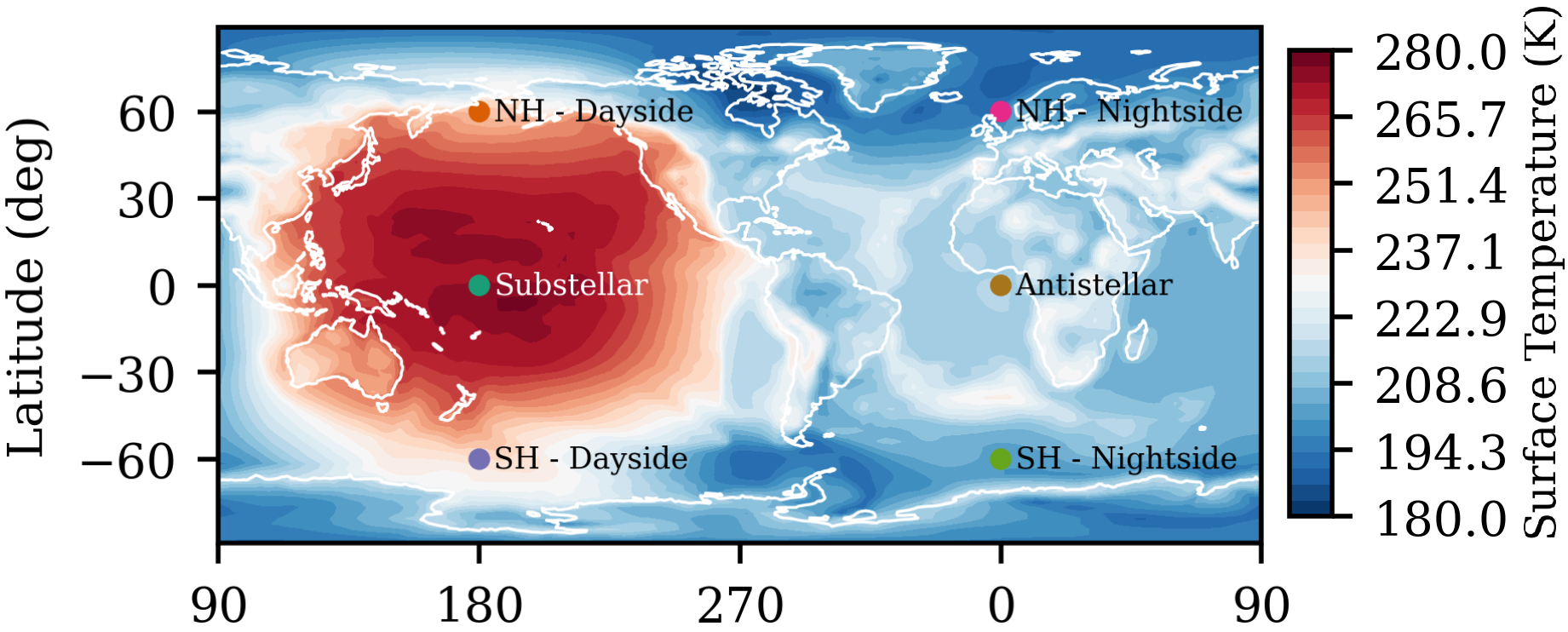}
\caption{ The mean surface temperature of our reference tidally-locked planet as a function of latitude and longitude. Here we show the boundaries of our model's Earth-like land-ocean distribution using white contours and the impact locations of each of our six cometary impacts as labelled, coloured, points. 
\label{fig:impact_location}  }
\end{centering}
\end{figure*}

TRAPPIST-1e is a terrestrial planet which remains a significant object of interest in the ongoing search for a potentially habitable exoplanet. It is slightly smaller than the Earth ($\mathrm{R} = 0.91\,\mathrm{R_\Earth}$ and $\mathrm{M} = 0.772\,\mathrm{M_\Earth}$) and orbits significantly closer to its cool host M-dwarf star ($\mathrm{P} \simeq 6.08\,\mathrm{days}$). As a result, the orbital and rotational periods of TRAPPIST-1e have likely synchronised, tidally-locking the planet such that the same side always faces TRAPPIST-1. And, since TRAPPIST-1 is a cool M-dwarf, its low luminosity means that, even in this short orbit, TRAPPIST-1e lies within the habitable zone, with a peak insolation of $900\,\mathrm{W\,m^{-2}}$, around $66\%$ that of the Earth. In our model, this insolation takes the form of a rescaled TRAPPIST-1 spectrum (taken from \citealt{2019ApJ...871..235P}) and centred over the Pacific Ocean (see the surface temperature in \autoref{fig:impact_location} which mirrors the insolation). 

As a result of this fixed insolation, the primary driver of large-scale atmospheric circulations on TRAPPIST-1e is the strong temperature contrast between the warm day-side and cold night-side (see \autoref{fig:impact_location}). This temperature contrast drives a global scale circulation with a strong upwelling on the day-side and downwellings on the night-side and over the poles.
On a planet without land (i.e., an ocean world), this circulation is symmetric between the northern and southern hemispheres (\citealt{10.1093/mnras/stad2704}); however in our simulations, both land-ocean boundaries, as well as the presence of orography (i.e., mountains) breaks this symmetry, leading to distinct flows in the northern and southern hemispheres (\citealt{sainsbury2024b}). 
In this work we will study how changing a comet's impact location within such an asymmetric global overturning circulation affects the strength of vertical advective mixing and hence diffusion-limited hydrogen escape. 
We consider six different impact locations, as marked on \autoref{fig:impact_location}: three on the day-side and three on the night-side, split between the equator and mid-latitudes ($\pm60^{\circ}$). The coupled impact models will be compared with the unimpacted reference atmosphere of \citetalias{2024arXiv240911151S}. 

\subsubsection{Exo-Earth-analogue} \label{ssub:exo_earth_analogue}

Our exo-Earth-analogue model is based upon the Earth's Pre-Industrial reference state that is distributed as part of the CESM 2.1 model release (\citealt{https://doi.org/10.1029/2019MS001882}). Both the short-wave and UV irradiation of the atmosphere are representative of the modern-day solar irradiation at the Earth's current orbital separation. The inclusion of a diurnal cycle means that, after the Coriolis effect, the main driver of atmospheric circulations is the temperature gradient between the equator and the poles. This results in smaller scale circulation cells which are highly efficient at mixing material horizontally, but much less efficient at driving material aloft (\citetalias{2025ApJ...990..117S}). As a consequence of this efficient horizontal mixing, we expect the impact location to have a relatively minor effect on the resulting vertical mixing and thus have chosen to focus on a single impact location for which data from an evolved coupled impact-climate model is already available (i.e., from \citetalias{2025ApJ...990..117S}). 

\subsection{Diffusion-Limited Hydrogen Escape} \label{sub:diffusion_limited_hydrogen_escape}

For an Earth-like planetary atmosphere, the diffusion-limited hydrogen escape rate, $\Phi_{\mathrm{escape}}$, is a function of the total mixing ratio of hydrogen-bearing molecules (\citealt{1974JAtS...31..305H,Catling_Kasting_2017}):
\begin{align}
	\Phi_{\mathrm{escape}} &= \sum_{i\in S} \frac{N_i \mathrm{f}(i) b_i }{H},
\end{align}
with
\begin{align}
	S&=\{\ce{H}, \ce{H2}, \ce{H2O}, \ce{OH}, \ce{CH4}\}.
\end{align}
Here, we consider a sum over all chemical species in the set $S$ where, for each species, $i$, $N_{i}$ is the number of hydrogen atoms per molecule, $\mathrm{f}(i)$ is the mixing ratio, and $b_i$ is the binary diffusion coefficient (see \autoref{eq:b5}). All these quantities, along with the atmospheric scale height $H$ are evaluated at a fixed pressure level, chosen to be representative of the turbopause which marks the switch from turbulent mixing of the atmospheric composition to compositional gradients which are set by molecular diffusion. On the Earth this corresponds with the region (at an altitude of $\sim 100$ km) in which the atmospheric temperature also starts to rapidly rise due to the decreased atmospheric density and strong UV irradiation. 

The binary diffusion coefficient quantifies the molecular mass transfer of each molecule through the background $\ce{N2}-\ce{O2}$ atmosphere, with a rate of mixing depending upon the local temperature of the atmosphere ($T$), and the mass and collisional cross-sectional area of each chemical species. The values for each hydrogen carrier are (\citealt{1973aero.book.....B}):
\begin{align}
	b_{\ce{H}} &= 6.5\times10^{17} T^{0.7},  \nonumber \\
	b_{\ce{H2}} &= 2.67\times10^{17} T^{0.75}, \nonumber \\
	b_{\ce{H2O}} &= 0.137\times10^{17} T^{1.072}, \nonumber \\
	b_{\ce{OH}} &= 0.137\times10^{17} T^{1.072},\,\mathrm{and}\nonumber \\
	b_{\ce{CH4}} &= 0.756\times10^{17} T^{0.747}. \label{eq:b5}
\end{align}
In CESM, both the hydrogen escape rates and mass-losses are post-processed with the upper boundary essentially acting as an infinite sink for any hydrogen-bearing molecule, thus ensuring that the abundances of hydrogen-bearing molecules at high altitudes (i.e. above the cold trap) is set by vertical advective and diffusive transport:  i.e the upper boundary condition ensures that our model remains in the diffusion limited regime. 

\section{Results} \label{sec:results}

An icy cometary impact with either of our tidally-locked or exo-Earth-analogue atmospheres results in significant water deposition which peaks in strength around $\sim30$ km above the surface. Here we investigate how this injection of water, and its associated changes on the planetary climate, affect the high altitude abundance of hydrogen and hydrogen-bearing molecules, and hence the diffusion-limited hydrogen escape rate. 

\subsection{Evolution of the Total Hydrogen Mixing Ratio} \label{sub:evolution_of_the_total_hydrogen_mixing_ratio}
\begin{figure*}[tp] %
\begin{centering}
\includegraphics[width=0.75\textwidth]{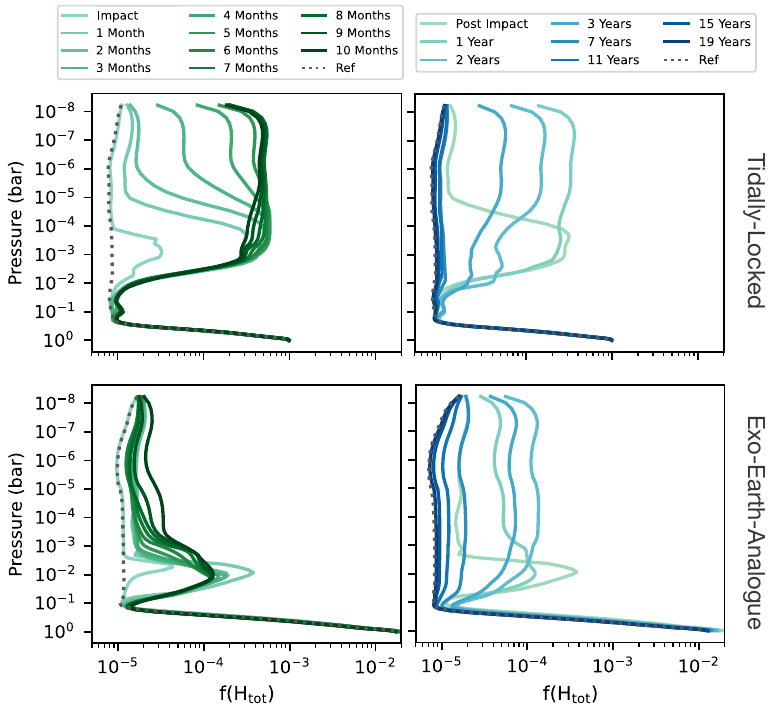}
\caption{ Monthly and global mean total hydrogen mixing ratio profiles ($\mathrm{f}(\mathrm{H}_\mathrm{tot})$) showing the enhancement in the high altitude abundance of hydrogen carrying molecules for an impact with the sub-stellar point of a tidally-locked atmosphere (top row) and over the Pacific Ocean of an Earth-analogue atmosphere (bottom row). Here we can see the rapid post-impact evolution of our tidally-locked atmosphere over the first 10 months post-impact (left/green) and the delayed enrichment of high-altitude hydrogen as well as the slow, but steady, settling of the atmosphere towards a quasi-steady-state (right/blue) reminiscent of the unimpacted reference state (grey dashed).  Note that the cometary impact occurs on the 26th of February in our model and hence appears weak in the `Impact' months mean profile (left-hand). As such, when we consider longer time-scale variations, we compare the evolved state with the mean atmosphere one month `Post Impact' (right-hand). 
\label{fig:F(H_Total)_evo}  }
\end{centering}
\end{figure*}
\begin{figure*}
\begin{centering}
\includegraphics[width=0.7\textwidth]{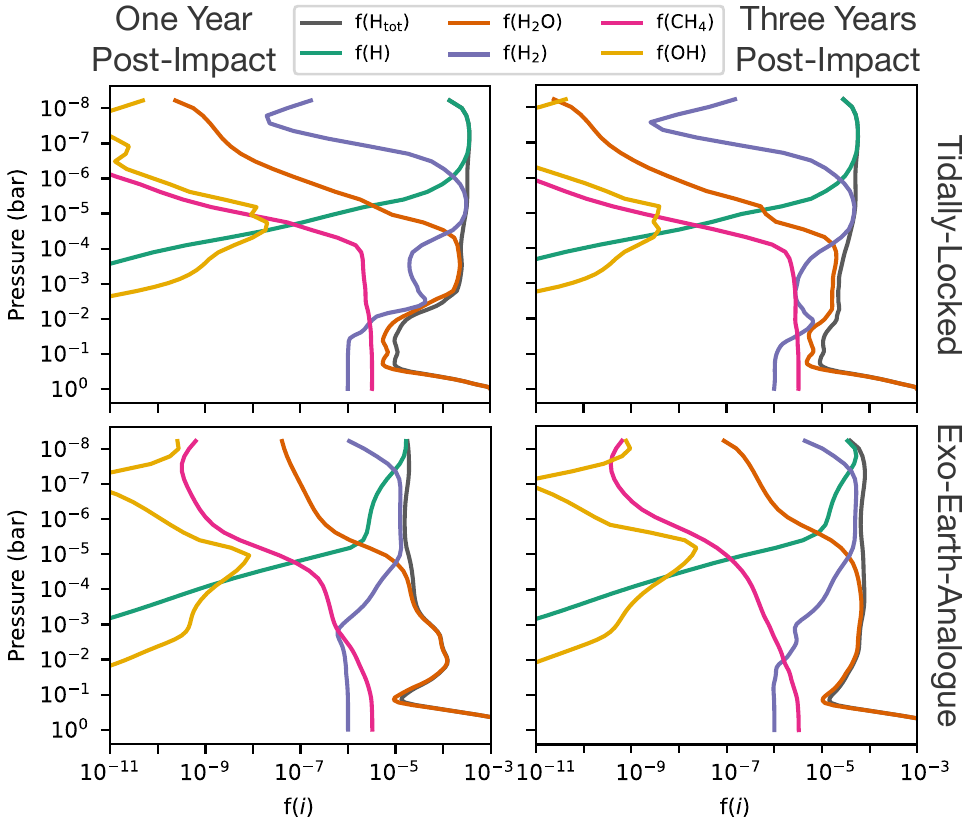}
\caption{Snapshots of the total hydrogen mixing ratio ($\mathrm{f}(\mathrm{H}_\mathrm{tot})$) and the mixing ratios of its constituents ($\mathrm{f}(\mathrm{H})$, $\mathrm{f}(\mathrm{\ce{H2O}})$, $\mathrm{f}(\mathrm{\ce{H2}})$, $\mathrm{f}(\mathrm{\ce{CH4}})$ and $\mathrm{f}(\mathrm{OH})$) for a comet impacting at the sub-stellar point of our tidally-locked, TRAPPIST-1e-like (top) atmosphere and over the Pacific Ocean in our Earth-analogue atmosphere (bottom). Note that we have selected these two points in time (one year post impact - left - and three years post impact - right) due to differences in the strength of vertical advective transport between tidally-locked and diurnally-rotating atmospheres. The online version of this figure includes an a movie showing the time evolution of the total hydrogen mixing ratio and its constituents.   \label{fig:f(H)_comp} }
\end{centering}
\end{figure*}
\begin{figure*}[tp] %
\begin{centering}
\includegraphics[width=0.75\textwidth]{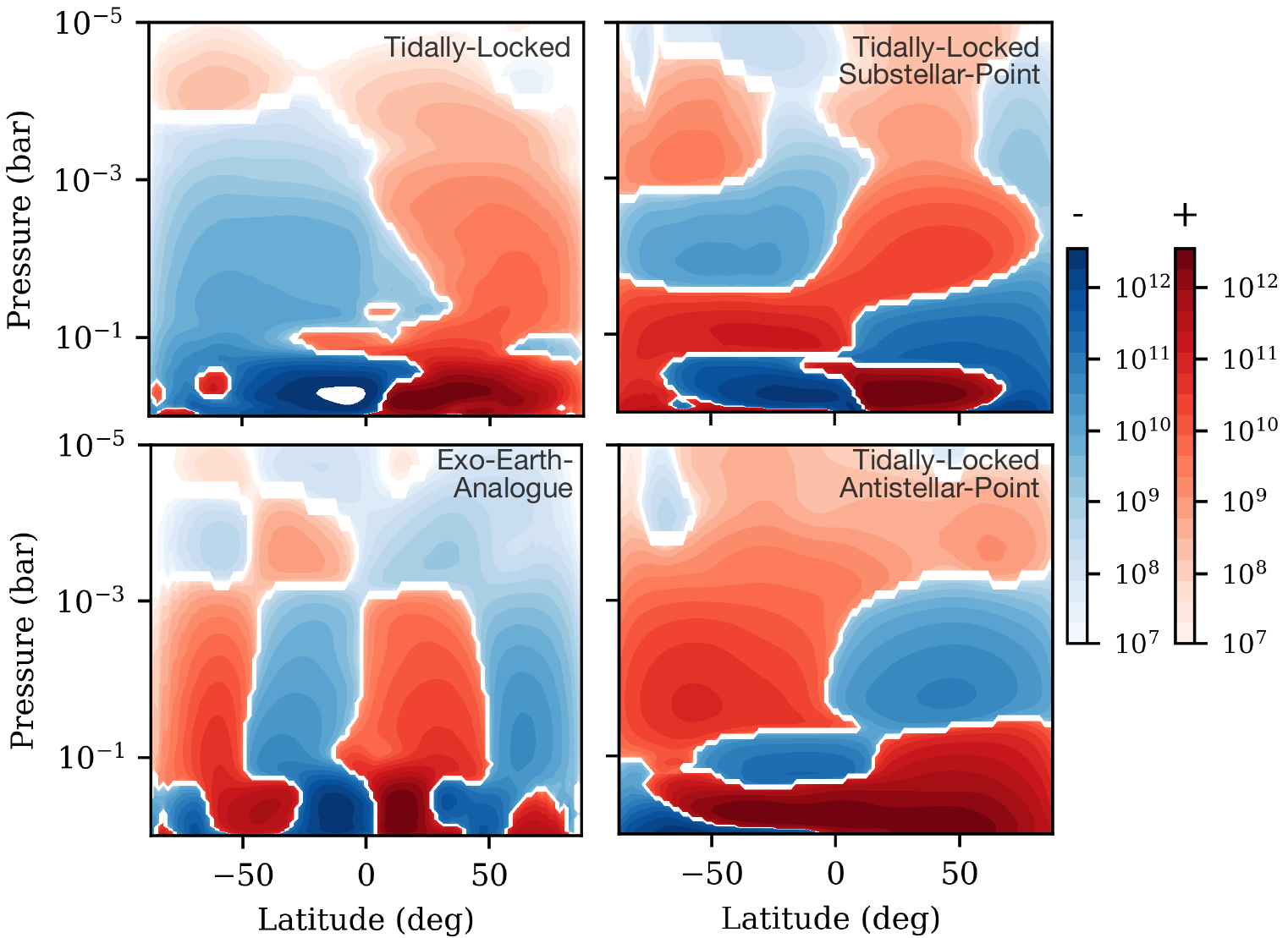}
\caption{ Select temporally averaged meridional mass streamfunctions taken from both tidally-locked (top row and bottom right) and exo-Earth-analogue (bottom left) atmospheric models. Panels in the left column show the (global) zonal mean of the circulation, whilst panels in the right column show the circulation calculated over a narrow ($15^\circ$ degree) zonal mean at the sub-stellar and anti-stellar points of our tidally-locked atmosphere respectively. Note that the meridional circulation is plotted on a log-scale with clockwise circulations shown in red and anti-clockwise circulations shown in blue. Together these circulations can combine to drive a net flow. 
\label{fig:streamfunction_comp}  }
\end{centering}
\end{figure*}

Initially, we focus our analysis on the coupled impact models of \citetalias{2024arXiv240911151S} and \citetalias{2025ApJ...990..117S}, investigating how the differences in vertical advection strength between a tidally-locked and diurnally-rotating (Earth-analogue) atmosphere affect the post-impact mixing ratio of hydrogen-bearing molecules ($\mathrm{f}(i)$). 

\autoref{fig:F(H_Total)_evo} plots snapshots of the evolution of the total hydrogen mixing ratio $\left(\mathrm{f}(\mathrm{H}_\mathrm{tot})=\sum_{i=S} \mathrm{f}(i)\right)$ over both the first 10 months post-impact (left) and the approximately twenty years required for the model to reach a quasi-steady-state (right). Note that we consider the model to have reached a quasi-steady-state when the rate of change of the fractional water abundance reaches a minimum. 

Comparing the time evolution of $\mathrm{f}(\mathrm{H}_\mathrm{tot})$ between models in which a comet impacts at the sub-stellar point of a tidally-locked atmosphere (top) or over the Pacific Ocean of an exo-Earth-analogue (bottom), the difference in vertical advection strength is immediately apparent. \\
In the tidally-locked atmosphere, within a month of the cometary impact the peak total hydrogen mixing ratio for pressures $<10^{-2}$ bar ($h\gtrsim 28$ km) has increased by a factor of $\sim30$. Further vertical transport both strengthens this enhancement, leading to a $\sim60$ times increase around six months post-impact, and drives high altitude mixing resulting in a near uniform enrichment for all pressures $<10^{-3}$ bar ($h\gtrsim 42$ km).
At very low pressures ($<10^{-5}$ bar - $\gtrsim70$ km), where escape can occur, an order of magnitude enhancement in $\mathrm{f}(\mathrm{H}_\mathrm{tot})$ persists for at least two years post-impact. However, as a result of downwards transport associated with the global overturning circulation as well as the freeze out of water vapour at high altitudes (driving the formation of cloud ice: \citetalias{2024arXiv240911151S}), the enhancement in the total hydrogen mixing ratio drops to less than a factor of two four years post-impact. From here this weak enrichment is relatively persistent, with $\mathrm{f}(\mathrm{H}_\mathrm{tot})$ remaining enhanced relative to the non-impact reference state even at quasi-steady-state. This long-lasting enrichment is possible because the cometary material has delivered hydrogen-rich material, increasing the global abundance of hydrogen-bearing molecules.
Since the hydrogen escape rate is directly linked to the hydrogen mixing ratio, this suggests that the escape rate should be strongest in the first two years post-impact, tapering off as the simulation progresses. We confirm that this is the case in \autoref{sub:the_hydrogen_escape_rate}. 

Moving onto our model of an impact over the Pacific Ocean of an exo-Earth-analogue atmosphere (bottom panels of \autoref{fig:F(H_Total)_evo}), we find that the time evolution of the total hydrogen mixing ratio is different.
In particular, the enrichment of $\mathrm{f}(\mathrm{H}_\mathrm{tot})$ at low pressures (high altitudes) is significantly delayed when compared to the tidally-locked case. For example, when the enrichment of high altitude hydrogen is near its peak in the tidally-locked case, $\sim10$ months post-impact, in our exo-Earth-analogue atmosphere we find only a factor of three to four enhancement in $\mathrm{f}(\mathrm{H}_\mathrm{tot})$ for $P\lesssim10^{-3}$ bar ($h\gtrsim47$ km). On the other hand, whilst the high-altitude enrichment in the tidally-locked atmosphere has already started to drop two years post-impact, in our exo-Earth-analogue atmosphere it has only just reached its peak in the upper atmosphere with a factor of 15 increase compared to the un-impacted reference state. This enrichment is also longer lasting than that found in the tidally-locked scenario. 
Comparing the evolution towards quasi-steady-state shown on the right-hand column of \autoref{fig:F(H_Total)_evo}, the enhancement in $\mathrm{f}(\mathrm{H}_\mathrm{tot})$ for $P<10^{-1}$ bar ($h\gtrsim15$ km) in the exo-Earth-analogue case remains greater than a factor of two until at least seven years post-impact, and stronger than the contemporary enhancement found in the tidally-locked case until ten years post-impact. 

This suggests that, compared to the tidally-locked case, the peak in the hydrogen escape rate for our exo-Earth-analogue impact model should be both delayed and weaker. However the sustained low-pressure/high-altitude enrichment found here will likely lead to it having a longer tail.

\subsection{Vertical Distribution of Hydrogen-Bearing Molecules} \label{sub:vertical_distribution_of_hydrogen}

Whilst the enhancement in the total hydrogen mixing ratio appears to be somewhat uniform at pressures $<10^{-3}$ bar ($h\gtrsim 42$ km) for both scenarios, the distribution of the mixing ratios of the hydrogen-bearing molecules which make up $\mathrm{f}(\mathrm{H}_\mathrm{tot})$ is altitude dependent. In \autoref{fig:f(H)_comp}, we plot profiles of $\mathrm{f}(\mathrm{H}_\mathrm{tot})$ and each of its constituent components for both of the previously discussed impact models at two points in time: one year post-impact, when the enhancement in $\mathrm{f}(\mathrm{H}_\mathrm{tot})$ peaks for the tidally-locked scenario, (top) and three year post-impact, around the peak of $\mathrm{f}(\mathrm{H}_\mathrm{tot})$ enhancement for the exo-Earth-analogue atmosphere. 

Starting with the tidally-locked (sub-stellar point) case both post-impact, and at quasi-steady-state, there are three dominant carriers of hydrogen in the atmosphere: water (\ce{H2O}), molecular hydrogen (\ce{H2}), and atomic hydrogen (\ce{H}). At quasi-steady-state, water is the dominant carrier of hydrogen for all pressures $>10^{-4}$ bar ($h\lesssim 56$ km). Above this altitude range, and on the day-side, water experiences ongoing UV irradiation and photodissociates, forming both hydrogen (H) and hydroxyl radicals (OH), as demonstrated by the peak in hydroxyl radical abundance around this level. The relatively high densities of hydrogen, hydroxyl radicals (\ce{OH}),  and hydroperoxyl radials (\ce{HO2}) in the photodissociation region mean that species rapidly combine to form molecular hydrogen (\ce{H2}), leading to \ce{H2} being the primary carrier of hydrogen between $\sim 10^{-4}$ and $\sim 10^{-6}$ bar. Finally, as we approach the top of our atmospheric model ($\lesssim10^{-6}$ bar - $\gtrsim86$ km), high energy particles (i.e. the solar wind) and UV photons drive the dissociation of molecular hydrogen leaving atomic hydrogen as the primary hydrogen carrier.
This distribution remains the same even after a cometary impact. Most of the impact-delivered water is deposited at $P>5\times10^{-2}$ bar ($h\sim 32$ km), where water is already the dominant carrier of hydrogen (top-left  of \autoref{fig:f(H)_comp}). It is then rapidly carried aloft to high altitudes, where it photodissociates, forming molecular and atomic hydrogen at low pressures where it can readily escape. 

The distribution for the exo-Earth-analogue case is very similar (bottom row of \autoref{fig:f(H)_comp}), except here differences in the pressure-temperature profile, driven by the diurnal heating of the Earth, shift the pressures at which the dominant hydrogen-bearing species changes. At quasi-steady-state the dominant carrier is water (vapour) for $P>10^{-5}$ bar ($h\lesssim77$ km), molecular hydrogen between $10^{-5}$ and $10^{-7}$ bar, and atomic hydrogen for $P<10^{-7}$ bar ($h\gtrsim105$ km). These changes are once again driven by the photodissociation of water and the dissociation of molecular hydrogen. As is the case for the tidally-locked regime, since most of the impact-delivered water is deposited deep into the atmosphere ($\sim10^{-2}$ bar - \citetalias{2025ApJ...990..117S}), the dominant carriers remain unchanged post-impact. Note that the peak in the material deposition profile occurs above the cold trap (which can be seen at $\sim0.1$ bar on the bottom-right panel of \autoref{fig:f(H)_comp}), suggesting that the primary control on the post-impact escape rate will be vertical advection/mixing. 

\subsection{Meridional Mixing and Vertical Advection} \label{sub:meridional_mixing_and_vertical_advection}

The differences in post-impact vertical advection rates between our tidally-locked and exo-Earth-analogue impact cases can be best understood via the meridional mass streamfunction $\psi$, which describes the transport of mass on the meridional plane (i.e., the latitude-pressure plane). This takes the form:
\begin{equation}
	\psi = \frac{2\pi R_{p}}{g\cos\theta}\int_{P_{top}}^{P_o}vdP,
\end{equation}
where $v$ is the latitudinal velocity, $R_{p}$ is the radius of the planet, $g$ is the surface gravity, $\theta$ is the latitude, and $P_{o}$ and $P_{top}$ are the pressure at the surface and top of the atmosphere respectively.

The left-hand column of \autoref{fig:streamfunction_comp} shows the zonally and temporally averaged meridional stream function for our tidally-locked (top) and exo-Earth-analogue (bottom) atmospheres. Here clockwise circulations are shown in red and anti-clockwise circulations are shown in blue, and where these circulations meet a net flow develops. Note that we do not plot the meridional mass streamfunction at very low-pressures due to the low local density of this region making mass flows very weak even on a log-scale. However in both rotation regimes we find a single circulation cell in each hemisphere which alternates between clockwise and anti-clockwise circulation with altitude, driving efficient mixing. 

From \autoref{fig:streamfunction_comp} the difference in circulation regimes between a diurnally-rotating and tidally-locked atmosphere is immediately apparent.
On a diurnally rotating world, after Coriolis forces, the main driver of atmospheric circulations is the equator to pole temperature gradient. This drives a multi-celled circulation profile (bottom-left), like that found in the Earth's atmosphere. Near the surface we find three pairs of circulation cells: Hadley cells near the equator, Ferrel cells at mid latitudes, and polar cells over the poles. Above this, the Brewer-Dobson circulation drives efficient meridional mixing which extends to $P\sim10^{-3}$ bar ($h\sim47$ km). At the equator and over the poles this drives a net upflow, whereas at the tropics, where the Hadley and Ferrel cells meet, we find a net downflow. As discussed in \citetalias{2025ApJ...990..117S}, these circulations are highly efficient at mixing material horizontally. Consequently, we do not anticipate that the hydrogen escape rate for our exo-Earth-analogue would be highly sensitive to the impact location, particularly for the range of impact latitudes-longitudes we consider for the tidally locked case (below).   
 Note, however, that these circulations are much less efficient at advecting material to $P<10^{-3}$ bar ($\gtrsim 47$ km), thus explaining the delayed and weakened enrichment of high-altitude hydrogen-bearing molecules (\autoref{sub:evolution_of_the_total_hydrogen_mixing_ratio}).  

On the other hand, on a tidally-locked world, the primary driver of atmospheric circulations is the strong, permanent, day-night temperature gradient. This drives global scale winds which manifest as a net equatorial upflow in the mean meridional circulation (top left) extending to $P<5\times10^{-5}$ bar ($h\gtrsim 60$ km). However the global nature of this wind means that we see a significant difference in the circulations between the day-side (top right), where we find the upwelling part of the global overturning circulation, and the night-side (bottom right), where we find down-flows (see \citealt{sainsbury2024b} for a more detailed discussion of these differences). As we discuss below, this makes the hydrogen escape rate highly sensitive to impact location on a tidally-locked world. Note that this sensitivity is further compounded by the asymmetry in land-mass distribution between the northern and southern hemispheres which, for example, shifts the zonal-mean equatorial upflow slightly into the northern hemisphere.  

We can see examples of these differences if we compare how impact-delivered material, the majority of which is deposited between $10^{-2}$ and $10^{-3}$ bar (around $32$ km in altitude), will interact with the circulations on the day and night-side. On the day-side, an equatorial impact leads to material being deposited in the upflow, rapidly carrying it aloft. However, for an off-equator impact, material will need to be carried to the equator before ascending. In the southern hemisphere this will be relatively efficient since the anti-clockwise circulation will carry material directly to the equator. However in the northern hemisphere this process will be slower as the clockwise circulation cell first advects water into the southern hemisphere before eventually carrying it towards to equator. 

Moving to the night-side, the global overturning circulation means that, for material to be carried aloft, it first needs to be advected to the day-side by near-surface winds. Hence even an equatorial impact will drive a delayed spike in high-altitude hydrogen bearing molecules. As for off-equator impacts, in the northern hemisphere the localised anti-clockwise circulation somewhat efficiently carries material to the equator whereas the extended clockwise circulation in the southern hemisphere will first carry material to low-pressures/high-altitudes where it will then be caught in the global overturning circulation and be advected towards the day-side via the surface.

\subsection{The Hydrogen Escape Rate} \label{sub:the_hydrogen_escape_rate}
\begin{figure*}[tp] %
\begin{centering}
\includegraphics[width=0.85\textwidth]{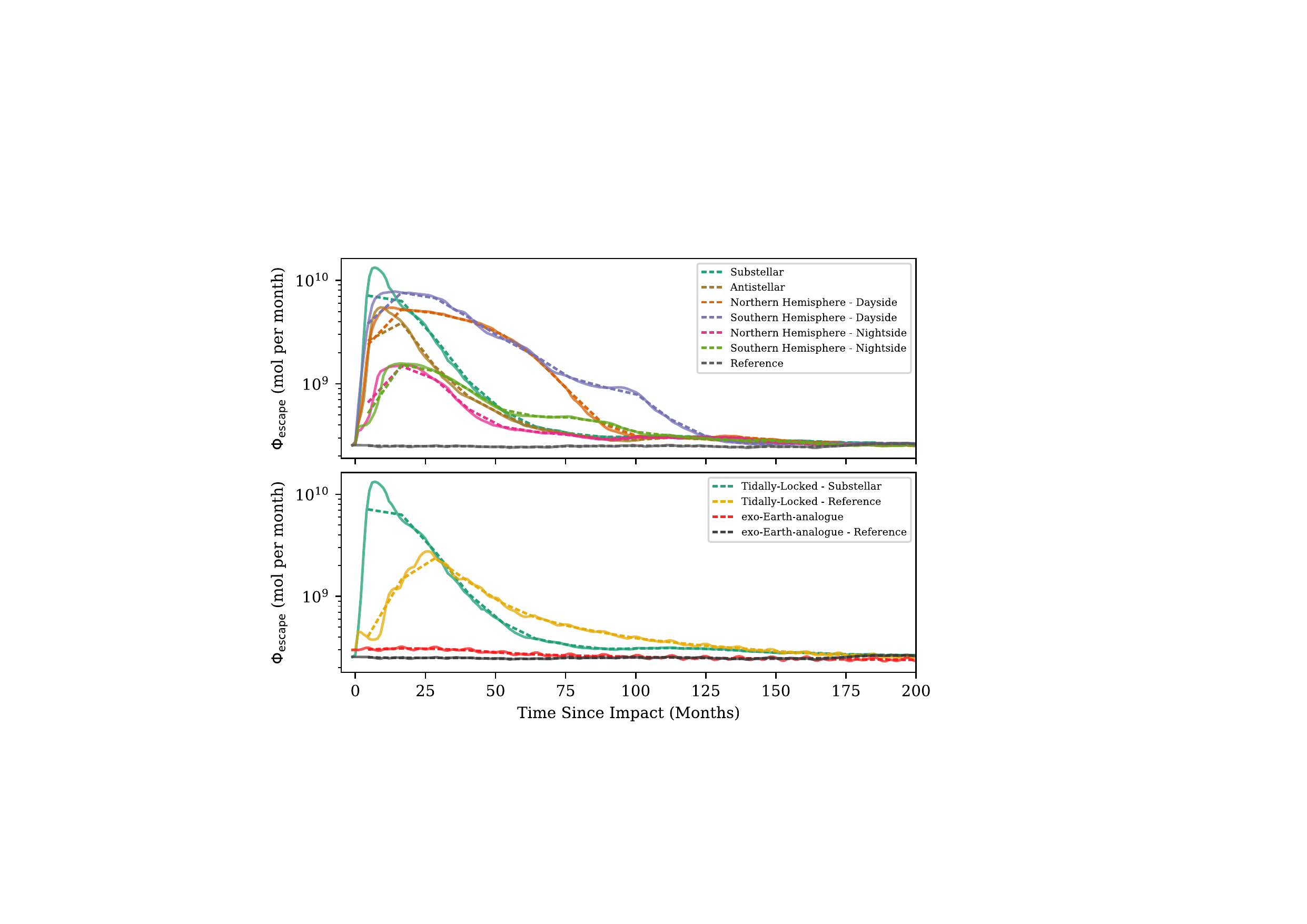}
\caption{ Time evolution of the annual mean (dashed lines) and monthly mean (fainter solid lines) total hydrogen escape rate ($\Phi_{\mathrm{escape}}$ [mol mth$^{-1}$]) for each of our seven different models of a cometary impact models. The top panel shows a comparison of the six difference models of a cometary impact with a tidally locked atmosphere whilst the bottom panel compares an impact with the sub-stellar point of a tidally-locked atmosphere with an impact over the Pacific Ocean of an exo-Earth-analogue atmosphere. We also plot the escape rate from our unimpacted tidally-locked (grey) and exo-Earth-analogue (red) reference states.  Note that the oscillations in the exo-Earth-analogue cases are driven by the inherent seasonality of the model. 
\label{fig:Phi_comp}   }
\end{centering}
\end{figure*}

The top panel of \autoref{fig:Phi_comp} shows the time evolution of the hydrogen escape rate ($\Phi_{\mathrm{escape}}$) for all six tidally-locked impact models considered here, as well as for the non-impacted reference case (grey, with a mean escape rate of $\Phi_{\mathrm{ref}}=2.5\times10^{8}\,\mathrm{mol\,mth^{-1}}$). The peak escape rates for each impact model can also be found in \autoref{tab:mass_loss}. Note that, for all cases discussed here, the impact-delivered water fully settles out of the atmosphere within twenty years of the impact, leading to escape rates that have fully converged with the non-impacted reference state.  \\

In general we find that an icy cometary impact drives at minimum an order of magnitude increase in the hydrogen escape rate in the first two to eight years post-impact with the location of material deposition within the global-scale atmospheric circulation setting both the strength and timing of this enhancement.

The strongest and longest lasting enhancements in the hydrogen escape rate are found for day-side impacts. For example, an impact at the sub-stellar point drives a two order of magnitude ($\Phi_{\mathrm{escape}}=1.33\times10^{10}\,\mathrm{mol\,mth^{-1}}$) enhancement in the escape rate only eight months post-impact. However since the cometary material is deposited directly in the global overturning circulation this enhancement is relatively short lived as material is also rapidly carried away from the day-side. Moving off-equator, we find a delayed, peaking 15 months post-impact, and weaker, $\Phi^{\mathrm{SH}}_{\mathrm{escape}}=7.8\times10^{9}\,\mathrm{mol\,mth^{-1}}$/$\Phi^{\mathrm{NH}}_{\mathrm{escape}}=5.41\times10^{10}\,\mathrm{mol\,mth^{-1}}$ enhancement in the hydrogen escape rate with a longer tail driven by a more spread-out delivery of water to the sub-stellar point (see \autoref{sub:meridional_mixing_and_vertical_advection}).

Moving to the night-side, whilst an impact at the anti-stellar point drives an enhancement in the hydrogen-escape rate that is similar in peak magnitude ($\Phi_{\mathrm{escape}}=5.47\times10^{10}\,\mathrm{mol\,mth^{-1}}$ ten months post-impact) to the off-equator impacts on the day-side, it is much shorter lived, with a lifetime closer to that of a sub-stellar impact. Finally, off-equator night-side impacts drive the weakest enhancement in the hydrogen escape rate ($\Phi^{\mathrm{SH}}_{\mathrm{escape}}=1.57\times10^{9}\,\mathrm{mol\,mth^{-1}}$ and $\Phi^{\mathrm{NH}}_{\mathrm{escape}}=1.51\times10^{9}\,\mathrm{mol\,mth^{-1}}$). These off-equator impacts also lack the tail of their day-side equivalents. This occurs because water is not a passive component of the atmosphere, it can evolve via both chemical reactions and precipitate out of the atmosphere as rain, snow, and cloud formation. This can include precipitating out of the atmosphere when passing through the cold trap during advection. As such, if the transport of water to the day-side upwelling is particularly delayed, or requires that water is advected towards the surface before reaching the sub-stellar upwelling, much of the deposited water vapour will evolve out of the atmosphere before being carried aloft.  

Given the importance of the underlying atmospheric circulations in setting the strength and lifetime of the enhancement in the hydrogen escape rate, how does the situation change when we consider a diurnally-rotating planet? \\
The lower panel of \autoref{fig:Phi_comp} shows the time evolution of $\Phi_{\mathrm{escape}}$ for both an impact with the sub-stellar point of our tidally-locked atmosphere (teal) and over the Pacific Ocean in our exo-Earth-analogue (yellow). Here the efficient horizontal mixing and inefficient vertical mixing found in an Earth-like atmosphere leads to a delayed, peaking around two and a half years post-impact,  enhancement in the hydrogen escape rate, peaking at $\Phi_{\mathrm{escape}}=2.76\times10^{9}\,\mathrm{mol\,mth^{-1}}$, significantly lower than the sub-stellar, tidally-locked, scenario.  
However, much like an off-equator day-side impact with a tidally-locked atmosphere, we find that this enhancement in the escape rate has a slightly extended tail when compared to an impact at the sub-stellar point. 

\subsection{Total and Excess Hydrogen Mass Loss} \label{sub:total_and_excess_hydrogen_mass_loss}
\begin{figure*}[tbp] %
\begin{centering}
\includegraphics[width=0.8\textwidth]{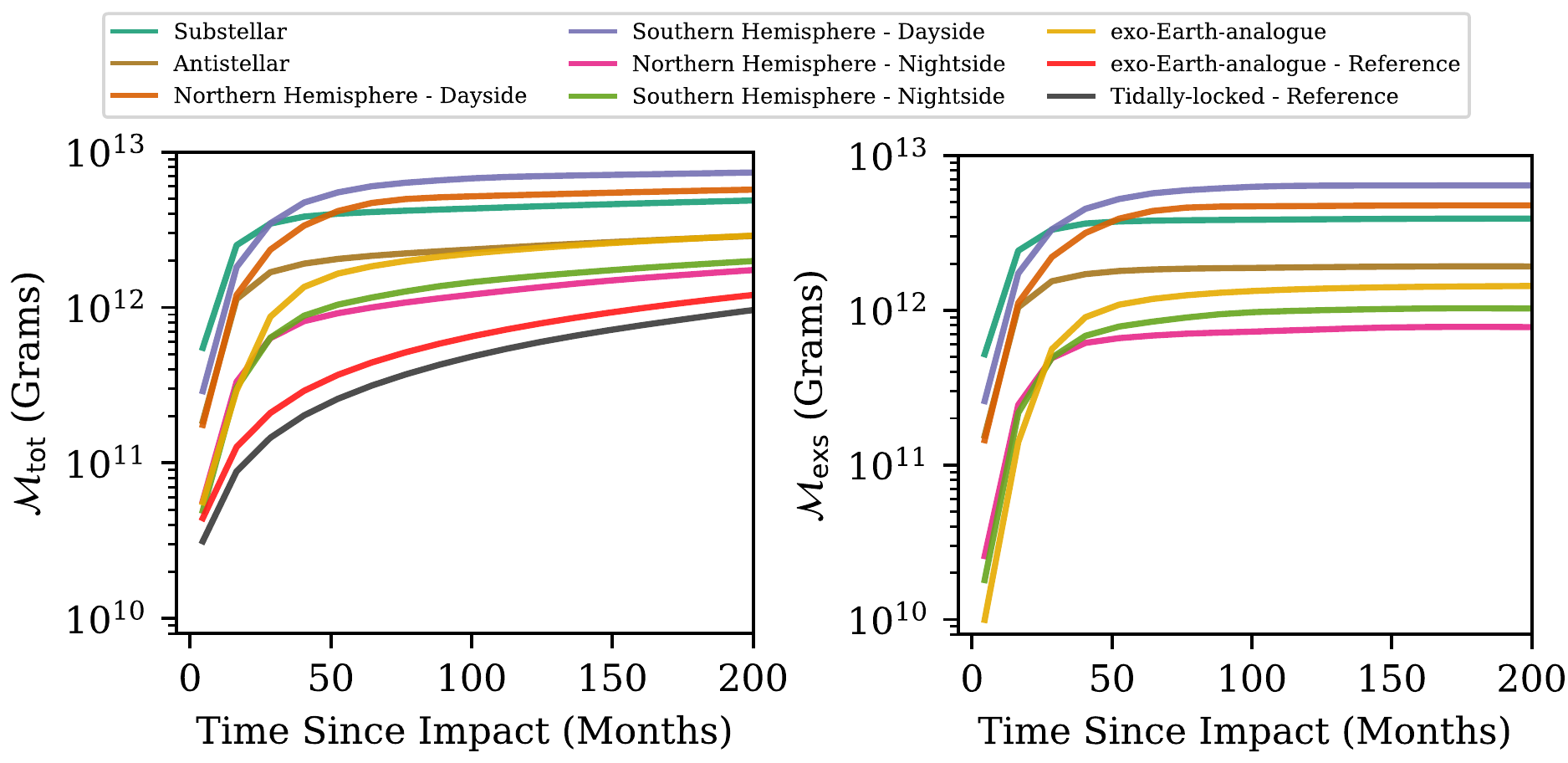}
\caption{ Time evolution of the cumulative total ($\mathcal{M}_{\mathrm{tot}}$ - left) and excess ($\mathcal{M}_{\mathrm{exs}}$ - right) hydrogen mass loss for each of our seven different models of a cometary impact with either a tidally-locked or exo-Earth-analogue  atmosphere. \label{fig:mass_loss}  }
\end{centering}
\end{figure*}

We next investigate how differences in the peak strength and overall lifetime of the enhancement in the hydrogen escape rate affect the total mass of hydrogen that would be lost.
In \autoref{fig:mass_loss} we plot the time evolution of the cumulative total ($\mathcal{M}_{\mathrm{tot}}$ - left) and excess ($\mathcal{M}_{\mathrm{exs}}$ - right) hydrogen mass loss for each of the seven impact models considered here, as well as the unimpacted reference states (grey), where appropriate.  Here we define: 
\begin{equation}
	\mathcal{M}_{\mathrm{exs}} = \mathcal{M}_{\mathrm{tot}} - \mathcal{M}^{\mathrm{ref}}_{\mathrm{tot}}. 
\end{equation}
Values for the total and excess hydrogen mass loss at quasi-steady-state are given in \autoref{tab:mass_loss}. 

In general, for an impact with a tidally-locked atmosphere we find that the trends in $\mathcal{M}_{\mathrm{tot}}$ and $\mathcal{M}_{\mathrm{exs}}$ shown in \autoref{fig:mass_loss} reflect our analysis of the hydrogen escape rate (\autoref{sub:the_hydrogen_escape_rate}).  Whilst an impact at the sub-stellar point drives the strongest post-impact enhancement in $\Phi_{\mathrm{escape}}$, which is reflected by a rapid rise in both $\mathcal{M}_{\mathrm{tot}}$ and $\mathcal{M}_{\mathrm{exs}}$, the sustained enhancement in $\Phi_{\mathrm{escape}}$ found in the off-equator day-side impacts drives a higher integrated hydrogen mass loss, up to $66\%$ higher for an impact in the southern hemisphere compared to the sub-stellar point. This occurs because water is delivered to the sub-stellar upwelling over a longer period of time, allowing for sustained mass loss.

However, when we move to the night-side the atmospheric processing of impact-delivered water on its journey to high altitudes means that the total and excess mass of hydrogen lost is significantly reduced. For instance we find that for an anti-stellar impact $\mathcal{M}_{\mathrm{exs}}$ is half of that for a sub-stellar impact, dropping to $26.1\%$ or $19.7\%$ of the sub-stellar value for an off-equator, night-side, southern/northern hemisphere impact respectively. 

As for an impact with a diurnally-rotating exo-Earth-analogue atmosphere, the increase in horizontal, and decrease in vertical, mixing efficiency (\autoref{sub:meridional_mixing_and_vertical_advection}/\citealt{2025ApJ...990..117S}) drives a total hydrogen mass loss which falls between the anti-stellar and night-side off-equator tidally locked scenarios. That is to say we find that $\mathcal{M}_{\mathrm{tot}}$/$\mathcal{M}_{\mathrm{exs}}$ is of the order of that found in the least efficient tidally-locked scenarios.

\begin{deluxetable*}{lccccc}[tb!]

\tablecaption{Peak hydrogen escape rate ($\Phi_{\mathrm{escape}}$) in $\mathrm{mol\,mth^{-1}}$ and $\mathrm{mol\,yr^{-1}}$, total hydrogen mass loss ($\mathcal{M}_{\mathrm{tot}}$) in grams, and excess hydrogen mass loss $\mathcal{M}_{\mathrm{exs}}$ in grams and as a $\%$ of the hydrogen mass deposited by the impacting comet.  \label{tab:mass_loss} }
\tablehead{Model & Peak $\Phi_{\mathrm{escape}}$ ($\mathrm{mol\,mth^{-1}}$) & Peak $\Phi_{\mathrm{escape}}$ ($\mathrm{mol\,yr^{-1}}$) & $\mathcal{M}_{\mathrm{tot}}$ (g) & $\mathcal{M}_{\mathrm{exs}}$ (g) & $\mathcal{M}_{\mathrm{exs}}$ (\% of initial hydrogen)}
\startdata
Sub-Stellar & $1.33\times10^{10}$ & $1.60\times10^{11}$ & $5.43\times10^{12}$ & $4.02\times10^{12}$  & 0.060 \\
SH - Day & $7.80\times10^{9}$ & $9.35\times10^{10}$ & $8.08\times10^{12}$ & $6.67\times10^{12}$ & 0.100 \\
NH - Day & $5.41\times10^{9}$ & $6.49\times10^{10}$ & $6.03\times10^{12}$ & $4.62\times10^{12}$  & 0.069 \\
Anti-Stellar & $5.47\times10^{9}$ & $6.56\times10^{10}$ & $3.26\times10^{12}$ & $1.85\times10^{12}$  & 0.028 \\
SH - Night & $1.57\times10^{9}$ & $1.89\times10^{10}$ & $2.46\times10^{12}$ & $1.05\times10^{12}$  & 0.016 \\
NH - Night & $1.51\times10^{9}$ & $1.81\times10^{10}$ & $2.20\times10^{12}$ & $7.90\times10^{11}$  & 0.012\\
TP1E Ref & $2.67\times10^{8}$ & $3.20\times10^{9}$ & $1.41\times10^{12}$ & -  & - \\
\hline
Earth & $2.76\times10^{9}$ & $3.31\times10^{10}$ & $2.93\times10^{12}$ & $1.51\times10^{12}$  & 0.023 \\
Earth Ref & $3.20\times10^{8}$ & $3.84\times10^{9}$ & $1.42\times10^{12}$ & -  & -
\enddata
\end{deluxetable*}

\subsection{Hydrogen Escape Mechanisms} \label{sub:escape_mechanisms}

In almost all of the impact scenarios discussed above, an icy cometary impact leads to at least an order of magnitude increase in $\Phi_{\mathrm{escape}}$ when compared with an un-impacted atmosphere. However, as seen in \autoref{fig:Phi_comp}, this enhancement is relatively short lived, being primarily driven by the photodissociation of vertically advected, impact-delivered, water. Water which, in addition to photodissociating, can also evolve out of the atmosphere both chemically and via precipitation and cloud-(ice) formation (\citetalias{2025ApJ...990..117S}).  Consequently we estimate that only a small-fraction of the impact-delivered hydrogen is lost via diffusion-limited escape mechanisms (\autoref{tab:mass_loss}), peaking at $0.1\%$ of the deposited cometary hydrogen for the case of a day-side impact on the southern hemisphere of a tidally-locked atmosphere. 

However this is only one mechanism via which a cometary impact may drive atmospheric escape. Impacts deliver a lot of energy to a small region of the atmosphere in a very short period of time, around $10^{21}$J for a $R=2.5$ km spherical comet made of pure water ice (\citetalias{2024arXiv240911151S}). This will drive localised heating (i.e. a shock\footnote{CESM and other Earth System Models are unable to model the effects of shocks. CESM in particular losses stability when temperature gradients are steep necessitating the slower and spread out thermal energy deposition considered here and in \citetalias{2024arXiv240911151S} and \citetalias{2025ApJ...990..117S}}) creating a plume of material with velocities that can exceed the escape velocity (\citealt{CAMERON1983195}). Such a plume can drive total atmospheric mass-loss of up to $10\%$ of the {\it total} mass of the impacting comet, although values are more typically on the order of $0.1\%$ to $1\%$ of the total impactor's mass (\citealt{10.1130/0-8137-2356-6.695,2009M&PS...44.1095S}). This suggests that shocks due to the passage of the impacting comet may drive mass loss rates that are $10\rightarrow1000$ times higher than diffusion-limited hydrogen escape as assumed here. However rather than being made up of hydrogen alone, the composition of this escaping material will be set by the column of atmosphere that the comet interacts with, i.e., mostly $\ce{N2}$ and $\ce{O2}$ with only a small fraction of the escaping material consisting of hydrogen.

As such, diffusion-limited escape remains one of the primary mechanisms via which impact-delivered hydrogen can escape the atmosphere, thus allowing for impacts to drive more heavy element enrichment than their compostion might imply.  This should remain true even if we consider more physically motivated cometary compositions (see the discussion below - \autoref{sub:assumptions_and_caveats}) in which hydrogen-rich water makes up $<100\%$ of the mass of the comet whilst remaining the primary constituent, i.e., $>50\%$. The effect may also become more significant when we consider an ongoing bombardment state in which the enhancement in high altitude water abundance can be effectively retained for longer periods of time. 

\subsection{Assumptions and Future Work} \label{sub:assumptions_and_caveats}

In this work we have explored the effects of pure water ice cometary impacts on an atmosphere which is representative of the modern-day Earth. However both the frequency of impacts as well as their relative effect on a terrestrial planets climate was likely higher earlier in a planets evolutionary history. Impacts would be more common since the number of objects in the planetary system would be higher, as remnants of the protoplanetary disc. Further, the system would be more dynamically excited due to effects such as planetary migration (as was the case in the solar system - \citealt{2023PhyU...66....2M}), increasing the likelihood that comets would be perturbed inwardly. 
For the Earth, this means that impacts were more likely earlier in its oxygenation history, when a lower oxygen (and ozone) abundance would lead to a rather different temperature-pressure profile, particularly in the stratosphere (\citealt{10.1098/rsos.211165}). This in turn will affect the temperature and location of the cold trap and hence the hydrogen escape rate. The lower oxygen content of the atmosphere, as well as the potential loss of impact-delivered hydrogen to space, also means that impacts have the potential to shape the global oxygenation state of the atmosphere. We will study the effects of impacts on the climate, hydrogen escape rates, and potential atmospheric oxygenation state (i.e. atmospheric metallicity) of Earth-like terrestiral worlds as part of a future study. 

Similarly, whilst the current coupled impact-atmosphere model assumes that the impacting comet is purely made of water-ice, observations and measurements of interstellar ices (e.g. \citealt{2023NatAs...7..431M}) and solar-system comets (e.g. \citealt{2017RSPTA.37560252B})  suggest otherwise. These observations suggest that comets should be dusty and that ices should contain a mix of volatiles including \ce{H2O}, \ce{CO}, \ce{CO2}, \ce{CH4}, and \ce{NH3}. In a future study, we will investigate how considering a more complex, and physically motivated, cometary composition affects the response of an Earth-like terrestrial atmosphere to both a single cometary impact and ongoing bombardment. This includes its potential effect on the hydrogen escape rate.  For example, a change in the tensile strength of the cometary ice due to a change in composition will effect the pressure-level at which breakup occurs. In turn, this will moidfy where water is primarily deposited in the atmosphere, potentially affecting the ability of mixing circulations to carry material aloft. 

Finally, whilst our model includes surface topography it is modelled after the modern-day Earth. As such, the differences between the hydrogen escape rates in the northern and southern hemisphere can be linked to differences in, for example, the land-fraction between the northern and southern hemisphere: much of the land is in the northern hemisphere for the Earth. On a terrestrial exoplanet the land-ocean distribution is likely to be rather different and as such our results should just be taken as indicating how the surface can affect physical processes, e.g. diffusion limited hydrogen escape, which are occurring far above.

\section{Concluding Remarks} \label{sec:concluding_remarks}

In this work we have investigated how the underlying planetary orbital and hence atmospheric circulation regime affects the vertical advection of hydrogen-bearing molecules, such as water, to low-pressures/high-altitudes where atmospheric escape of hydrogen becomes diffusion-limited. We have explored seven distinct impact scenarios: six cometary impacts with a tidally-locked atmosphere in order to quantify the role of the global overturning circulation in shaping the diffusion-limited escape rate and an equatorial impact over the Pacific Ocean of an exo-Earth-analogue chosen to demonstrate the effects of Earth-like atmospheric circulations on diffusion-limited escape. 
For each scenario we considered the impact of of a spherical comet of pure water-ice with $R=2.5\,\mathrm{km}$ and $\rho=1\,\mathrm{g\,cm^{-3}}$, evolving our atmosphere to a quasi-steady-state in which the rate of change of the fractional water abundance has reached a minimum. Our investigation of the evolution of the hydrogen escape rate in these seven different scenarios revealed the following key results:
\begin{itemize}
	\item On a tidally-locked planet, the primary carrier of hydrogen is water for $P>10^{-4}$ bar ($h\lesssim 56$ km), molecular hydrogen between $10^{-4}$ and $10^{-6}$ bar, and atomic hydrogen at $P<10^{-6}$ bar ($h\gtrsim86$ km). For our exo-Earth-analogue, the inclusion of diurnal heating modifies the atmospheric temperature-pressure profile such that water is the primary carrier for $P>10^{-5}$ bar ($\lesssim77$ km), molecular hydrogen dominates between $10^{-5}$ and $10^{-7}$, and atomic hydrogen can be found at the very top of the atmosphere ($P<10^{-7}$ bar - $h\gtrsim105$ km). 
	\item In all impact scenarios we found that the total mixing ratio of hydrogen-bearing molecules (i.e. \ce{H}, \ce{H2}, \ce{OH}, \ce{H2O}, and \ce{CH4}) increase by more than an order of magnitude at low pressures: $P\lesssim10^{-3}$ bar ($h\gtrsim 42$ km) for a tidally-locked atmosphere and $P\lesssim10^{-2}$ bar ($h\gtrsim 30$ km) for an exo-Earth-analogue. 
    \item This post-impact increase in the low pressure abundance of hydrogen-bearing molecules drives a corresponding increase in the hydrogen escape rate $\Phi_{\mathrm{escape}}$. In the tidally-locked regime, our models indicate that the largest increase in the peak hydrogen escape rate ($\Phi_{\mathrm{escape}}=1.33\times10^{10}\,\mathrm{mol\,mth^{-1}}$) is found for an impact at the anti-stellar point. However this peak is relatively short-lived and weaker, $\Phi_{\mathrm{escape}}=7.80\times10^{9}$ (NH) and $5.41\times10^{9}\,\mathrm{mol\,mth^{-1}}$ (SH),  but longer lasting increases in the diffusion-limited escape rate are found for day-side impacts in the southern/northern hemisphere respectively. This is reflected in the total mass loss, which increase from $\mathcal{M}_{\mathrm{exs}}=4.02\times10^{12}\,\mathrm{g}$ for a sub-stellar impact to $6.67\times10^{12}\,\mathrm{g}$ for a day-side impact in the southern hemisphere. 
    \item Significantly smaller enhancements in $\Phi_{\mathrm{escape}}$ are found for impacts on the night-side of our tidally-locked planet. We find a peak hydrogen escape rate of $5.47\times10^{9}\,\mathrm{mol\,mth^{-1}}$ for an impact at the anti-stellar point. We also find that the overall weakest enhancement in the escape rate occurs for a northern-hemisphere night-side impact: $\Phi_{\mathrm{escape}}=1.51\times10^{9}\,\mathrm{mol\,mth^{-1}}$. We find correspondingly reduced excess mass losses, $\mathcal{M}_{\mathrm{exs}}=1.85\times10^{12}\,\mathrm{g}$ or $\mathcal{M}_{\mathrm{exs}}=7.90\times10^{11}\,\mathrm{g}$ respectively. 
	\item Impacts with the Earth drive diffusion-limited escape rates which are closer to those found on the night-side rather than the day-side of a tidally-locked atmosphere. We find a peak hydrogen escape rate of $\Phi_{\mathrm{escape}}=2.7\times10^{9}\,\mathrm{mol\,mth^{-1}}$, which corresponds to an excess mass loss of $\mathcal{M}_{\mathrm{exs}}=1.51\times10^{12}\,\mathrm{g}$. 
\end{itemize}

The differences in escape rates between our seven different impact scenarios were driven by atmospheric circulations and the resulting efficiency of advecting impact-delivered material to higher altitudes. However these differences in circulation regime also affect the diffusion-limited hydrogen escape rate in our unimpacted reference atmospheres. Despite the atmosphere of our reference tidally-locked atmosphere being globally drier than its exo-Earth-analogue counterpart, we find very similar escape and mass-loss rates. 
%
%
This occurs because, even though the planet is cooler than the Earth and much of its surface is frozen, the day-side stellar insolation is strong enough to ensure that any surface oceans are liquid, thus providing a ready source of water which is immediately deposited in the sub-stellar upwelling. This ensures that it can be readily carried aloft to low pressure regions where it can undergo photolysis and drive diffusion limited escape. However the global atmosphere remains dry due to the divergent high altitude day-side winds carrying any remaining water vapour to the night-side where it can freeze/precipitate out of the atmosphere. A similar result can be found in the Sub-Stellar Point over Land (SSPL) case of \citet{sainsbury2024b} where only a small liquid ocean is found near the coast of the sub-stellar land-mass (Africa). Here we find $\Phi_{\mathrm{escape}}=2.2\times10^{8}\,\mathrm{mol\,mth^{-1}}$, around $10\%$ lower than the peak escape rate found in our reference tidally-locked case. This suggests that any water that evaporates from the ocean is rapidly carried away from the surface. 

Overall our analysis suggests that, even after a cometary impact, diffusion-limited escape remains the primary mechanism via which hydrogen can escape from the atmosphere, and its post-impact enhancement may have important implications for the oxygenation of planetary atmospheres (\citealt{CATLING20051}), increasing the amount of material that comets can deliver whilst maintaining the relatively high metallicity of Earth-like atmospheres.  

Finally we find that understanding the underlying planetary circulation regime is, amongst other things, a key component of determining the diffusion-limited escape rate. This reinforces the need for global, coupled, atmospheric models which can study the transport of key atmospheric constituents whilst including processes which can deplete them during their journey through the atmosphere.

\begin{acknowledgements}
\nolinenumbers
The authors would like to thank Daniel Marsh for valuable discussions when finalising the final version of this work. \\
F. Sainsbury-Martinez and C. Walsh would like to thank UK Research and Innovation for support under grant numbers MR/T040726/1 and MR/Z00019X/1. Additionally, C. Walsh would like to thank the University of Leeds and the Science and Technology Facilities Council for financial support (ST/X001016/1).\\
This work used the DiRAC Data Intensive service (DIaL3 - project dp371) at the University of Leicester, managed by the University of Leicester Research Computing Service on behalf of the STFC DiRAC HPC Facility (\url{www.dirac.ac.uk}). The DiRAC service at Leicester was funded by BEIS, UKRI and STFC capital funding and STFC operations grants. DiRAC is part of the UKRI Digital Research Infrastructure.
\end{acknowledgements}

\bibliography{papers}{}
\bibliographystyle{aasjournal}

\end{document}